# Plasmonic modulator based on gain-assisted metal-semiconductor-metal waveguide

Viktoriia E. Babicheva[1,2,*], Irina V. Kulkova[1], Radu Malureanu[1], Kresten Yvind[1], Andrei V. Lavrinenko[1]

[1]Department of Photonics Engineering, Technical University of Denmark, Ørsteds Plads, Bld. 343, DK-2800 Kongens Lyngby, Denmark
[2]Moscow Institute of Physics and Technology, Institutsky Pereulok 9, 141700 Dolgoprudny, Russia

* Corresponding author. Tel.: +45 45256883; fax: +45 4593 6581.
E-mail address: vbab@fotonik.dtu.dk (V.E. Babicheva)

**Abstract.** We investigate plasmonic modulators with a gain material to be implemented as ultra-compact and ultra-fast active nanodevices in photonic integrated circuits. We analyze metal-semiconductor-metal (MSM) waveguides with InGaAsP-based active material layers as ultra-compact plasmonic modulators. The modulation is achieved by changing the gain of the core that results in different transmittance through the waveguides. A MSM waveguide enables high field localization and therefore high modulation speed. Bulk semiconductor, quantum wells and quantum dots, arranged in either horizontal or vertical layout, are considered as the core of the MSM waveguide. Dependences on the waveguide core size and gain values of various active materials are studied. The designs consider also practical aspects like n- and p-doped layers and barriers in order to obtain results as close to reality. The effective propagation constants in the MSM waveguides are calculated numerically. Their changes in the switching process are considered as a figure of merit. We show that a MSM waveguide with electrical current control of the gain incorporates compactness and deep modulation along with a reasonable level of transmittance.

## 1. Introduction

Nanoplasmonics is a booming focus area in nanophotonics with potential applications in sensing and integrated circuits [1, 2]. However, until now the main obstacle in the implementation of plasmonic nanocomponents is the presence of high losses in connection with highly-confined plasmonic waves propagation. Alternatives to the conventional noble metals, silver and gold, are desirable and have been proposed recently [3-6].

A way to compensate losses is the introduction of gain materials [7-9]. Recently, loss compensation by gain in hybrid dielectric loaded plasmonic waveguides has been studied [10, 11]. Garcia-Blanco et. al. [10] analyzed long-range dielectric-loaded surface plasmon-polariton waveguides and loss compensation by the rare-earth-doped double tungstate crystalline material as a gain medium. A hybrid plasmonic waveguide with a nano-slot was examined in [11], where also potential gain media were discussed.

Nevertheless, hybrid dielectric loaded plasmonic waveguides suffer from optical confinement (device sizes are of the order of microns). The strongest mode confinement can be achieved in metal-insulator-metal (MIM) waveguides [12, 13]. The MIM waveguide is a promising structure. It provides a base for extremely fast and efficient ultra-compact plasmonic devices, including modulators [14, 15], photodetectors [16] and lasers [17]. However, there is a trade-off on applications: the MIM waveguide allows high fields localization, but by contrast, propagation losses in the waveguide are also high.

Loss compensation by a gain semiconductor core in MIM waveguides led to the concept of metal-semiconductor-metal (MSM) waveguide. Gain-assisted propagation in a subwavelength MSM waveguide was theoretically analyzed [9]. The thickness of the active core of high-index semiconductor was varied from 50 nm to 500 nm. For a gold MSM waveguide the required critical gain coefficient for a lossless propagation is achievable at telecommunication wavelengths. More recent systematic theoretical investigation of giant modal gain and amplified propagation in MSM structures with silver plates and the 100-200 nm semiconductor cores is presented in [18]. The optimum gain level, which gives the strongest plasmonic resonance and slowing down of the group velocity is demonstrated.

Encapsulation of InP-based semiconductors in metal and configuring compact MSM waveguides was discussed regarding lasing application. Lasing in metallic-coated nanocavities, namely in the MSM waveguide with rectangular cross-section InP/InGaAs/InP pillars is realized [19-21]. Theoretical analysis of a MSM waveguide with the InGaAs core [22] shows the possibility of realization of a plasmonic semiconductor nano-laser. Semiconductor electrons-holes dynamics in conduction and valence bands is described numerically. The influence of intermediate low index buffer layers in a MSM sandwich is studied as well.

One of the hottest topics is active plasmonics, which combines semiconductor electronics and nonlinear optics to control the optical properties of different nanodevices [23, 24]. Recently an all-semiconductor active plasmonic system based on InAs heterostructures has been proposed for mid-infrared operation wavelengths [25]. An InGaAsP quantum wells stack was tried to improve properties of negative index materials [26, 27]. Whereas the goal of the work was to design the fast optical modulation (tens picoseconds time scale) in the Si spacer layer of such metamaterials, gain produced rather low changes in transmission because of metal screening.

From another point of view, changing the permittivity and thus controlling the propagation in a MIM waveguide is an attractive subject of active plasmonics. A straightforward way is to vary the transmission through the waveguide by changing the permittivity of a sandwiched medium. This idea led to theoretical description of switches using gain-assisted MIM structures [28, 29], where the proposed solutions utilize a quantum-dots-doped semiconductor. It is also shown that Fabry-Perot resonances in the active core layer enhance the switching effect.

Another possibility to control propagation of light in a MIM waveguide by changing permittivity of the medium in side-coupled cavities was theoretically analyzed in [28, 30-32]. In particular, it was shown that implementation of gain media in parts of the waveguide can compensate losses along the device [30].

Despite the fact that photonic switches based on either bulk semiconductors or quantum dots have been studied, there is lacking of comprehensive research on MSM compact devices with the active core from well-defined materials. All proposed MSM structures with loss compensation and switching are analyzed in the assumption of some model gain material uniformly distributed along the dielectric core. Here we bring the subject closer to reality considering InGaAsP-based semiconductors in the MSM core.

We study the performance of a MSM device for plasmonic switching applications. In particular, ultra-compact modulator can be designed. The semiconductor core is considered consisting of a bulk gain medium, quantum wells or layers with quantum dots. We examine several potentially realizable designs of structures and analyze modulator's performance. In Section 2 we describe our theoretical model and numerical approach to MSM waveguide simulations. In Section 3 we analyze possible MSM waveguide arrangement. Performance of device with the bulk semiconductor core is studied in Section 4, with quantum-dots-based core in Section 5 and with quantum wells layers in Section 6. Discussion and final sum up of the results are in Section 7.

## 2. Simulation model

We performed frequency domain simulations using the commercial software package CST Microwave Studio [33]. The material gain $g$ is connected with the imaginary part of the semiconductor permittivity: $\varepsilon'' = -gn'/k_0$, where $k_0$ is the free-space wave-number and $n'$ is the real part of the refractive index. So, for positive gain values the imaginary part of permittivity is negative. Due to the CST constrains, materials with the negative imaginary part of permittivity cannot be implemented directly. To circumvent this constrain and perform CST numerical simulations with gain materials we follow the suggestion from [34]. It consists in editing the permittivity values, in particular changing the sign of $\varepsilon''$, in the Visual Basic script of the history list file. CST simulations were compared with analytical calculations to validate this approach.

Fabry-Perot resonances on a finite-length semiconductor core can significantly increase the effect of the total transmission change in a MSM waveguide (see e.g. [28] and [29]). Partial removing of lossy material from waveguide's core can increase transmittances as well [35, 36]. However, fabrication of a patterned core inside MSM is challenging and we do not consider it here. So we imply that plasmonic waveguides are uniform in the propagation direction and study device characteristics independently from its length.

We are interested in transmission properties of MSM structures at the telecom wavelength 1.55 μm. A MSM waveguide supports a surface plasmon polariton (SPP) wave in the transverse magnetic (TM) polarization. The silver plates (Fig.1) have the thickness of 120 nm, which is enough to keep the domain-termination error at negligible level. The silver permittivity is approximated by the Drude formula:

$$\varepsilon_{Ag} = \varepsilon_\infty - \frac{\omega_{pl}^2}{\omega^2 + i\gamma\omega}, \qquad (1)$$

where $\varepsilon_\infty = 1$, plasma frequency $\omega_{pl} = 1.38 \cdot 10^{16}$ s$^{-1}$ and collision frequency $\gamma = 3.22 \cdot 10^{13}$ s$^{-1}$. It gives $\varepsilon_{Ag} = -128.7 + 3.44i$ at $\lambda = 1.55$ μm.

Signal attenuation (extinction) per device's unit length can be defined as

$$A_{/L} = 10\lg(P_0/P)/L = 8.68 \cdot \text{Im}(k_{eff}), \qquad (2)$$

while logarithmic extinction ratio (ER) per unit length is

$$\text{ER}_{/L} = 10\lg(P_{on}/P_{off})/L = 8.68 \cdot \left(\text{Im}(k_{eff})_{off} - \text{Im}(k_{eff})_{on}\right), \qquad (3)$$

where $\text{Im}(k_{eff})_{on}$ and $\text{Im}(k_{eff})_{off}$ are the imaginary parts of the effective propagation constant of a waveguide mode in the on-state (voltage or electrical current switched on) and off-state (voltage or current switched off). ER shows how strong we can vary the mode propagation through the waveguide.

However, for plasmonic switching devices that possess significant propagation losses, another length independent figure of merit (FoM) can be defined (similar to [15, 35]):

$$\text{FoM} = \frac{\text{ER}_{/L}}{A_{/L}} = \frac{\left|\text{Im}(k_{eff})_{on} - \text{Im}(k_{eff})_{off}\right|}{\text{Im}(k_{eff})_{state}}, \quad (4)$$

where denominator is either $\text{Im}(k_{eff})_{on}$ or $\text{Im}(k_{eff})_{off}$ depending on which state is transmitting. FoM (4) describes how strong we can vary the mode propagation through the waveguide in comparison with the attenuation in transmitting state. ER or FoM should be chosen depending on particular waveguide properties. The high gain core in a MSM device can completely compensate plasmonic losses and cause wave amplification. In case of $\text{Im}(k_{eff})_{on} \cong 0$ we can receive infinitely high FoM (4) that does not have physical meaning. Most of structures studied in the paper allow complete loss compensation. Therefore, characterizing MSM devices we present results for $\text{ER}_{/L}$ (3). Spontaneous emission noise is not included in our model.

The on-state refers to a MSM waveguide, whose core (an InGaAsP-based semiconductor layer) is active and exhibits gain. Such active material can be designed to have a band gap at the wavelength of interest 1.55 μm. In the off-state the core material exhibits high losses interpreted as negative material gain. Zero-current negative gain is equal to the maximum material gain that can be theoretically achieved in the semiconductor material [37].

We also compared switching to an "insulator-state" of a MSM waveguide, which refers to the waveguide with a passive sandwiched medium. The passive medium means that the medium has a band gap at wavelengths shorter than the wavelength of interest (1.55 μm). So precisely at 1.55 μm it is transparent, i.e. it has neither gain nor loss caused by electron-hole recombination. Plasmonic related losses exceed those of the semiconductor core by far. So, in the "insulator-state", the MSM device has losses caused by the silver electrodes only.

Depending on the active region structure, i.e. bulk semiconductor, quantum wells (QWs) or quantum dots (QDs), and carrier density the spectral range of gain of the InGaAsP-based material can vary from tens to hundreds of nanometers around the telecom wavelengths. This characteristic influences the device operation bandwidth. The energy band gap of the gain medium is controlled by the fraction of each component in the $\text{In}_x\text{Ga}_{1-x}\text{As}_y\text{P}_{1-y}$ material, and therefore defines the wavelength of the maximum gain. Material should also be engineered based on operating carrier density. Components with higher bandgap, which is closer to wavelength of interest 1.55 μm, should be used for lower carrier density and consequently achievable gain (e.g. $\text{In}_{0.558}\text{Ga}_{0.442}\text{As}_{0.95}\text{P}_{0.05}$ has bandgap ~1.58 μm). However, we took data from [37] and studied bulk gain material $\text{In}_{0.53}\text{Ga}_{0.47}\text{As}$ that has bandgap ~1.69 μm and requires a high carrier density to achieve high gain at 1.55 μm.

The infrared refractive index for different $\text{In}_x\text{Ga}_{1-x}\text{As}_y\text{P}_{1-y}$ structures is approximately $n´ = 3.1+0.46y$ [38] and varies only slightly with respect to the wavelength. Therefore, we neglected the material dispersion.

In order to simulate an as-realistic-as-possible system we included the n- and p-doped layers at the top and bottom of the active layer (see Fig. 1b) to prevent the non-equilibrium distributions and recombination of carriers at the contacts. The n- and p-doped layers should have somewhat different composition to obtain the band gap at higher frequency. However, in calculations we fix the real part of permittivity $\varepsilon´ = 12.46$ (which corresponds to the real part of refractive index $n´ = 3.4...3.5$ for a small gain) for all studied semiconductors apart from permittivity of the InP layer fixed as $\varepsilon_{\text{InP}} = 10$, which corresponds to $n_{\text{InP}} = 3.16$. The imaginary part of effective propagation constant $\text{Im}[k_{\text{eff}}]$ and modal gain $g_\text{m}$ in the waveguide are proportional to each other: $g_\text{m} = -2\text{Im}[k_{\text{eff}}]$.

## 3. MSM waveguide arrangements

Realization of MSM waveguides based devices encounters several problems and can be accomplished in different ways. One way is to deposit metal plates as horizontal layers (Fig. 1a) and most of the fabricated devices are based on this procedure (e.g. [13, 14, 16]). However, InGaAsP-based heterostructures can be deposited only on a semiconductor native substrate, what means that some additional techniques, e.g. bonding or membrane etching and coating, should be applied to encapsulate semiconductor layers in metal. In such horizontal design n- and p-doped layers should be deposited between metal plates and gain medium while metal plates serves as electrodes. The horizontal layout is preferable for systems when a very narrow gap between metal plates is required. As a MSM waveguide can have width about $w \sim 1$ μm (see Fig. 1a), borders influence can be neglected, and the system can be simulated as two dimensional with $w$ taken as infinitely long. However, as the noble metals exhibit poor adhesion to III-V semiconductor materials this possibility may prove to be a challenge to fabricate without using e.g. Ti or Cr adhesion layers. Such lossy layers can significantly affect device operation.

The other option is to etch narrow ridges and to coat them with metal. This technique is applied in the case of fabrication of nanolaser structures [20, 22]. In such vertical design, the high aspect ratio (that is $H/d$, see Fig. 1c) is challenging to achieve in practice, despite the InP-based materials are relatively strong because of their crystalline structure. The high aspect ratio is required due to the mode confinement necessities. Because of finite aspect ratio $H/d$ the mode profile in such waveguide deviates from the two-dimensional case. Two cases can be considered here (Fig. 1c and d). The structure shown in Fig. 1c with very high aspect ratio was realized for lasing by Hill et. al. [20]: the core thickness was $d = 100...350$ nm, while the whole ridge height $H > 1$ μm. In Fig. 1c n- and p-doped layers are horizontal and can be deposited during the growth of the gain medium. In this case, metal plates do not serve as electrodes anymore and passivation layers (e.g. $Si_3N_4$ or InP) on both sides of the ridge are required. Optical or electrical pump should be applied in the vertical direction. That assumes large distance of current propagation and can slow down the modulation. Electrically driven fast modulation can be realized in the design of Fig. 1d, which requires n- and p-doped layers depositing after ridge etching. However, it is very challenging for practical realization.

We performed numerical simulations of a MSM waveguide of finite height H and various core thicknesses d (correspond to Fig.1c in the passive state). Fig. 2 shows propagation length L (L=1/ Im[$k_{eff}$]) for different waveguides in comparison with the ideal case (infinitely high waveguide). The waveguide is sandwiched between high-index material, the same as the core ($\varepsilon= 12.46$); metal corners are rounded for simulation reason with radius of curvature 4 nm. Even for H=1 μm the propagation length halves because of mode spreading outside waveguide (see inset of Fig. 2).

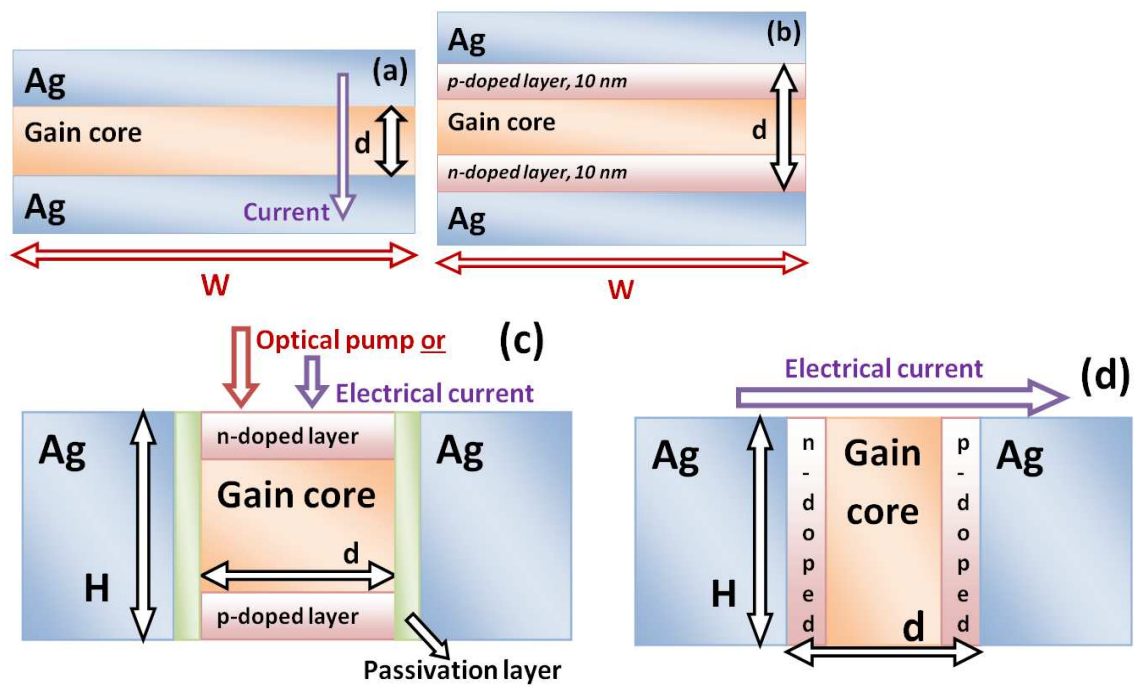

Fig. 1. Cross sections of MSM waveguides with the gain core. a) Basic MSM waveguide. Horizontal arrangement. b) Refined structure of the MSM waveguide which includes the gain core and n- and p-doped layers. Horizontal arrangement. c) Vertical arrangement with top pumping. d) Vertical arrangement with metal plates as electrodes.

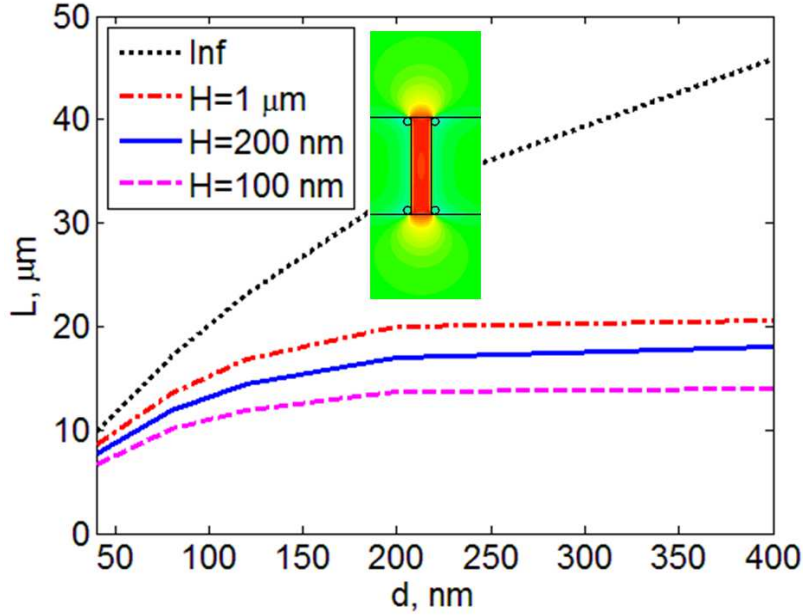

Fig. 2. Propagation length for MSM waveguide of finite height H in passive state. Inset: Electric field distribution for waveguide's cross section with H=200 nm and d=40 nm.

## 4. MSM waveguide with bulk gain medium

In the first approximation the semiconductor core can be considered as a layer of a uniformly distributed gain material (Fig. 1a). For the bulk gain core we set material gain $g_b$ in the range from $-0.6 \cdot 10^4$ cm$^{-1}$ to $0.6 \cdot 10^4$ cm$^{-1}$ for In$_{0.53}$Ga$_{0.47}$As composition at the telecom wavelength (equivalent to $\varepsilon'' = 0.5225...-0.5225$) with carrier density up to $6.7 \cdot 10^{19}$ cm$^{-3}$ [37]. The core thickness $d$ is varied from 20 nm to 400 nm. The main advantage of the double-sided metal-clad waveguide is the strong field localization between the plates. The inset in Fig. 3 shows the effective refractive index of the waveguide, growing from the refractive index of sandwiched material $n' = 3.53$ for a wide core up to 6.7 for a very thin core.

Fig. 3 shows Im[$k_{eff}$] versus the MSM core thickness for various gain values. We consider: the "insulator-state" for the passive semiconductor core, "off-state" for the maximum loss in semiconductor as well as several gain values. To validate the CST simulations with the negative imaginary part of permittivity we compare results with the ones obtained by analytical solution of the SPP dispersion equation for a three-layer system (Fig. 3). The results match perfectly.

As was discussed above the system has very high losses in the off-state. With the increase in current, the material gain becomes higher and, at some point, compensates losses. For example, for the gain value $g_{b1}= 0.1 \cdot 10^4$ cm$^{-1}$ ($\varepsilon''= -0.0871$) complete loss compensation in the MSM waveguide (Im[$k_{eff}$] = 0) can be achieved in the case of a core thickness $d \approx 60$ nm. Further increase of gain gives light amplification in such system. For gain values more than $g_{b2}= 0.34 \cdot 10^4$ cm$^{-1}$ ($\varepsilon''= -0.2961$) the strong field localization within the gap between metal plates starts to play a crucial role and the MSM design gives an essential benefit. In the case of high gain the Im[$k_{eff}$] dependency on the core thickness is monotonically increasing, opposite to the case of lower gain (see Fig.3). In other words, for high gain values the thinner the MSM core the longer propagation can be achieved.

In Fig. 4 we show Im[$k_{eff}$] in the MSM waveguide with a bulk gain medium for various carrier densities. For better comparison, we converted the material gain data [37] into the semiconductor permittivity and added in Fig 4 (dotted line). The imaginary part of semiconductor permittivity has nearly linear behaviour on the logarithmic scale as follows from carriers and photons rate equations, thus the imaginary part of the effective propagation constant exhibits the same tendency.

Further we consider an advanced structure, which includes n- and p-doped layers on both sides of the active layer (see Fig. 1b). We assume the n- and p-doped layers being 10 nm thick. These intermediate layers are approximated as insulator layers ($\varepsilon''= 0$) having the same permittivity as the semiconductor core, $\varepsilon'=12.46$. The partial removal of the gain material causes an increase in Im[$k_{eff}$]. In such MSM waveguides complete loss compensation with gain value $g_{b1} = 0.1 \cdot 10^4$ cm$^{-1}$ can be achieved for core thicknesses more than 100 nm (see Fig. 3).

Calculated ER of both systems versus the core thickness is shown in Fig. 5 for various transition states. Introduction of n- and p-doped layers for thin core waveguides decreases ER as expected.

As an additional option we also consider the difference between the on-state and passive state ("insulator-state") of the MSM waveguide. The "insulator-state", that is $g = 0$, can be achieved in the studied material for relatively low carrier density, namely $1.25 \cdot 10^{18}$ cm$^{-3}$ [37].

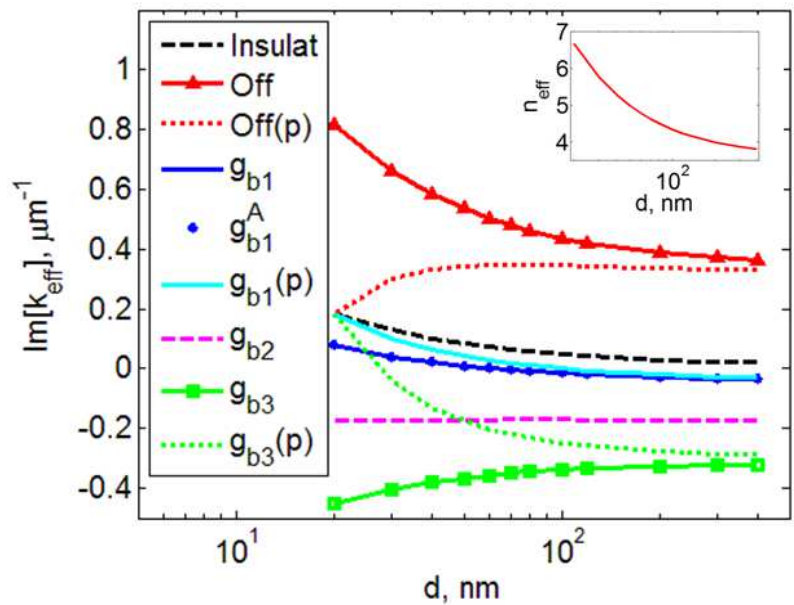

Fig. 3. Imaginary part of the effective propagation constant for various thicknesses of the bulk gain semiconductor core. All results are from CST numerical simulations apart from "$g_{b1}^A$" which is the analytical solution of the SPP dispersion relation for a three-layer system. Notation: "No" for passive waveguide (insulator-state); "Off" for switched off current $g_b = -0.6 \cdot 10^4$ cm$^{-1}$; "Off(p)" the same for refined structure with n- and p- doped layers; "$g_{b1}$" on-state with $g_{b1} = 0.1 \cdot 10^4$ cm$^{-1}$; "$g_{b1}^A$" for analytical calculation same as "$g_{b1}$"; "$g_{b1}(p)$" on-state with $g_{b1} = 0.1 \cdot 10^4$ cm$^{-1}$ in refined structure with n- and p- doped layers; "$g_{b2}$" on-state with $g_{b2} = 0.34 \cdot 10^4$ cm$^{-1}$; "$g_{b3}$" on-state with $g_{b3} = 0.6 \cdot 10^4$ cm$^{-1}$; "$g_{b3}(p)$" on-state with $g_{b3} = 0.1 \cdot 10^4$ cm$^{-1}$ and refined structure with n- and p- doped layers. Inset: Effective refractive index in the MSM waveguide filled with In$_{0.53}$Ga$_{0.47}$As bulk gain medium.

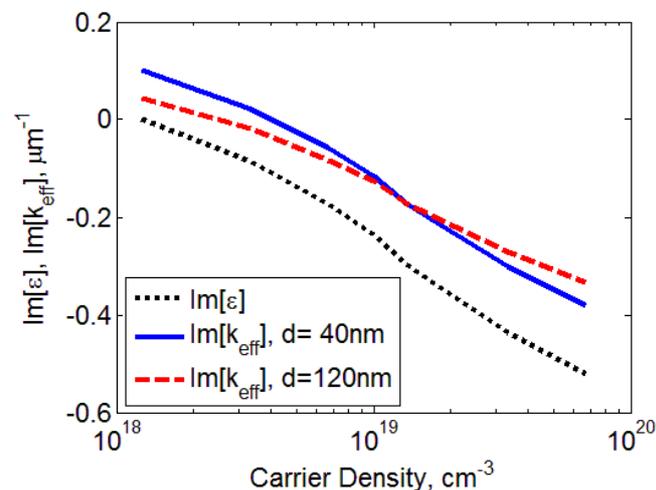

Fig. 4. Imaginary part of the effective propagation constant in the MSM waveguide with the bulk gain medium for various carrier densities. Imaginary part of semiconductor permittivity is shown for comparison.

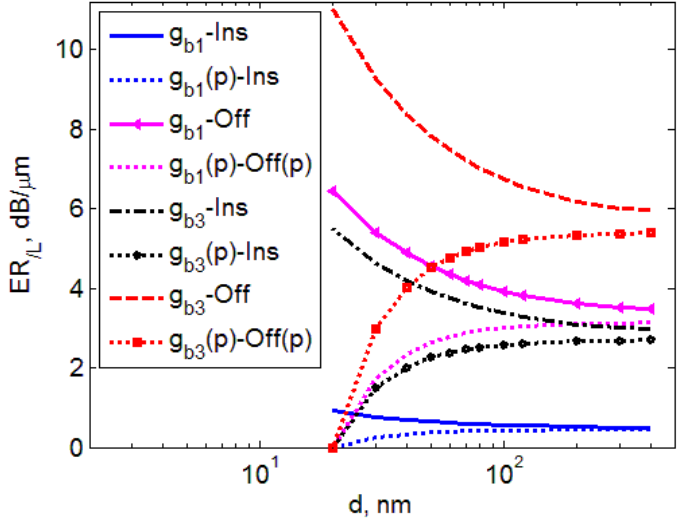

Fig. 5. ER of MSM device with the bulk gain core. Notation: "$g_{bi}$-Ins" shows switch between insulator-state with $g_b = 0$ and on-state with $g_{bi}$ ($g_{b1} = 0.1 \cdot 10^4$ cm$^{-1}$ and $g_{b3} = 0.6 \cdot 10^4$ cm$^{-1}$); "$g_{bi}$(p)-Ins" the same, but for the refined structure; "$g_{bi}$-Off" shows switch between off-state with $g_b = -0.6 \cdot 10^4$ cm$^{-1}$ and on-state with $g_{bi}$ ($g_{b1} = 0.1 \cdot 10^4$ cm$^{-1}$ and $g_{b3} = 0.6 \cdot 10^4$ cm$^{-1}$); "$g_{bi}$(p)-Off(p)" the same, but for the refined structure.

## 5. MSM waveguide with quantum dots

The second realistic option for the core design in MSM structures can be an InGaAsP layer with InAs QDs. Significant progress in QDs devices has been done at 1.3 μm wavelength [39, 40] while efficient emission on 1.55 μm wavelength is still under investigation [41, 42].

Despite on the small volume, QDs exhibit material gain more than one order of magnitude larger than QWs under the same injection current density because of the strong confinement and quantization of energy levels [43, 44]. However, such gain can be obtained only when the electric field is in the QDs plane. Thus, in the horizontal design (Fig. 6a) QDs layer gives quite low gain, e.g. material gain $g_{TM} = 0.4 \cdot 10^3 ... 1.2 \cdot 10^3$ cm$^{-1}$ in conversion of the QDs to an equivalent bulk layer [45]. Such low bulk gain we discussed in the previous section. Therefore depositing of QDs in horizontal design does not provide any additional benefits.

Further we consider the system in Fig. 6b. The regular vertical arrangement of QDs appears in Stranski–Krastanov growth [46] as well as grown by metalorganic vapour-phase epitaxy of columnar QDs (vertically aligned, closely stacked QDs) [45].

Parameters for numerical simulations of QDs are the following. A single QD has an ellipsoid shape; the in-plane cross-section is circular with 30 nm diameter, while the shortest ellipsoid axis is 5 nm. QDs are arranged in a regular array with period 50 nm in both in-plane directions. The described geometrical arrangement corresponds to $4 \cdot 10^{10}$ cm$^{-2}$ QDs density [41]. An InGaAsP matrix containing one QDs layer has thickness $s = 10$ nm and $\varepsilon' = 12.46$, the same as in the previous analysis. We performed simulation for a single QDs layer with periodic boundary conditions in vertical directions, that equivalent to the two-dimensional waveguide with an infinite number of QDs layers. We used $g_{d1} = 1 \cdot 10^4$ cm$^{-1}$ and $g_{d2} = 5 \cdot 10^4$ cm$^{-1}$.

The QDs volume ratio in the 10-nm thick stack layer is approximately 9%. For simulations we consider the equivalent system with bulk semiconductor (see Fig. 1a) having the gain value defined by the volumetric ratio and gain of the QDs. Thus, the average gain is $b_{d1} = 0.9 \cdot 10^3$ cm$^{-1}$ and $b_{d2} = 4.5 \cdot 10^3$ cm$^{-1}$.

The total thickness of semiconductor between the metal plates $d = 50 \cdot N$ nm, where $N = 1...5$ is a number of columns. Up to 5 columns containing QDs were considered for simulations. The imaginary part of the effective propagation constant and ER of MSM devices with QDs are shown in Fig. 7 and Fig. 8, respectively. Results for QDs and an effective bulk medium are perfectly matched, so the averaging procedure is justified. We assume that the perfect periodic arrangement of QDs in simulations does not influence system propagation properties, which are the same as for a core with randomly distributed QDs. While to achieve $b_{d2} = 4.5 \cdot 10^3$ cm$^{-1}$ for bulk material requires a very high current, it is feasible for the QDs effective medium.

For the highest gain $g_{d2} = 5 \cdot 10^4$ cm$^{-1}$ in the on-state Im[$k_{eff}$] monotonically increasing with the number of columns, which causes decreasing of ER. Average gain $b_{d2} = 4.5 \cdot 10^3$ cm$^{-1}$ > $g_{b2}$, so we observe the same trend in Im[$k_{eff}$] vs. size dependence as for the bulk material (see Fig. 3).

However, as was discussed above, the vertical design has weak mode confinement. In practice, up to 10-20 QDs layers with the same properties can be fabricated. It means that with vertical spacing $s = 10$ nm the total height of the structure cannot exceed $H_A \sim 200$ nm that is not enough for mode localization. So apart from this $H_A \sim 200$ nm of gain region the sandwiched structure should be complemented to $H \sim 1$ μm by another high-index material. Similar case was realized in [20], where MSM waveguide has $H_A = 300$ nm of active gain region, while the whole ridge height $H > 1$ μm. It increases Im[$k_{eff}$] of the whole waveguide, therefore, any partial filling by active material increases propagation losses and suppresses efficiency of modulator.

Another solution is to increase vertical spacing $s$ and obtain relatively uniform distribution of gain medium along waveguide's cross section. However, in practice, it is hard to obtain vertical alignment at large spacing. Thus if we assume, in order to achieve $H_A \sim 1$ μm, spacing $s = 50$ nm, the average gain will fall down to $b_{d2} = 0.9 \cdot 10^3$ cm$^{-1}$.

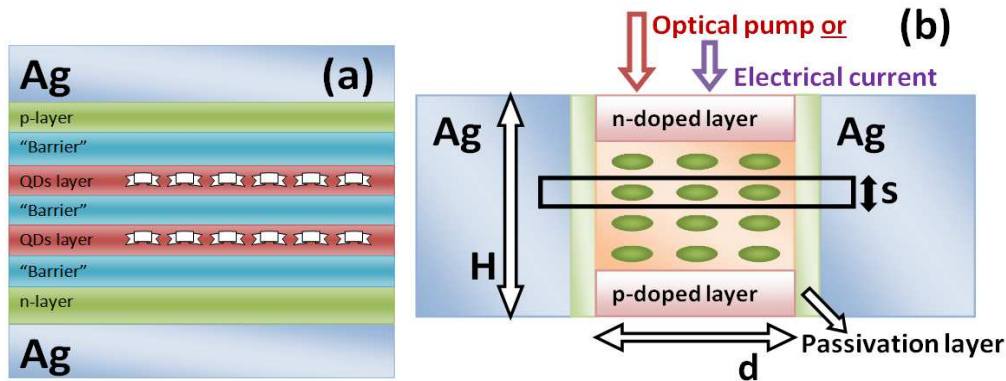

Fig. 6. Cross sections of the MSM waveguide with QDs layers. a) Horizontal design. b) Vertical design.

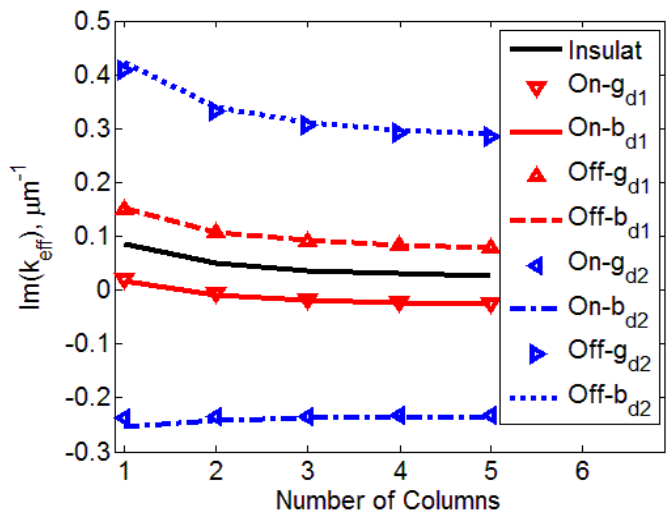

Fig. 7. Imaginary part of the effective propagation constant in the MSM waveguide with QDs. Notation: "Insulat" is passive waveguide, $g_{d1} = 1 \cdot 10^4$ cm$^{-1}$, $g_{d2} = 5 \cdot 10^4$ cm$^{-1}$, "$b_{d1}$" and "$b_{d2}$" corresponds to the MSM waveguide with uniform distribution of bulk gain material with $b_{d1} = 0.9 \cdot 10^3$ cm$^{-1}$ and $b_{d2} = 4.5 \cdot 10^3$ cm$^{-1}$.

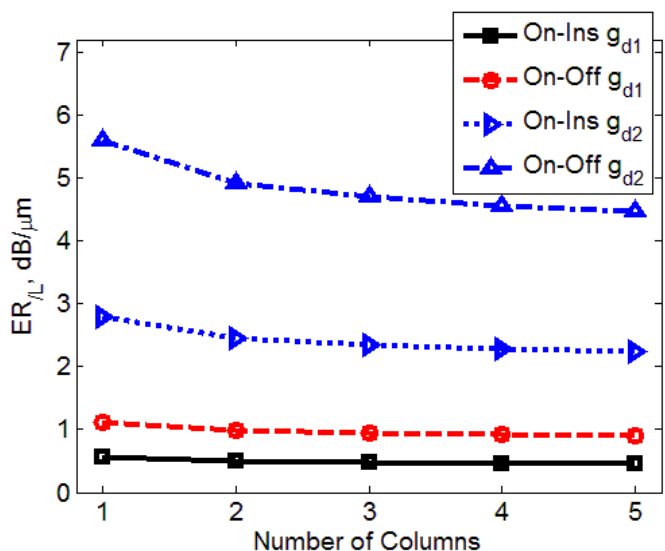

Fig. 8. ER of the MSM device with QDs. "On-Ins $g_{di}$" shows transition between insulator-state (g=0) and on-state with $g_{di}$. "On-Off $g_{di}$" shows transition between off-state with -$g_{di}$ and on-state with $g_{di}$ ($g_{d1}$= 1·10$^4$ cm$^{-1}$ and $g_{d2}$= 5·10$^4$ cm$^{-1}$).

## 6. MSM waveguide with quantum wells

As an alternative design to the MSM waveguide with a bulk semiconductor core we consider the MSM core, which consists of several QWs separated by barriers (Fig. 9a). Tensile strained quantum wells are used to ensure a strong interaction with the electrical field of the mode. We took In$_{0.466}$Ga$_{0.534}$As QWs and In$_{0.51}$Ga$_{0.49}$As$_{0.86}$P$_{0.14}$ barriers with thicknesses 5 nm. The composition is chosen such that to obtain the bandgap at 1.55 μm [47]. The wells are strain compensated which will allow stacking of many wells. However, due to the position of the barrier material in the miscibility gap we decided to limit the number of wells to 25. We performed simulations taking this number into account and defining the total core thickness based on it. In a QWs MSM system it is necessary to have at least one different medium with a higher energy band gap for the n- or p-doped layer. Therefore, we choose 10 nm-thick In$_{0.67}$Ga$_{0.23}$As$_{0.71}$P$_{0.29}$ layer and 10 nm-thick InP layer. As it was discussed in [22], adding a thin layer with low refractive index in between the waveguide and the metallic plates decreases the net gain in the system. This is due to confining the field outside the gain material, despite of lowering the fields in the metallic layer.

The results of numerical simulations for Im[$k_{eff}$] and ER are shown in Fig. 10 and Fig. 11, respectively. We studied two gain levels, $g_{w1}$ = 0.4·10$^4$ cm$^{-1}$ and $g_{w2}$ = 1·10$^4$ cm$^{-1}$, which correspond to $\varepsilon''$ = - 0.348 and $\varepsilon''$ = - 0.87 respectively. The former $g_{w1}$ is currently achievable, e.g. in tensile QWs [37, 48, 49], while the latter $g_{w2}$ is feasible [50]. In such MSM waveguides with QWs complete loss compensation appears either for more than three QWs with gain $g_{w1}$ or already with one QW with gain $g_{w2}$. In contrast to QDs system, for considered $g_{w1}$ and $g_{w2}$ values, ER of the QWs system is growing monotonously with the number of QWs and starting at some thicknesses (approximately d=90 nm) can be higher than in the case of the bulk semiconductor with $g_{b1}$= 0.1·10$^4$ cm$^{-1}$.

Vertical design can be considered as well (Fig. 9b). The estimation of structure efficiency can be conducted as following. One period of the structure has thickness $s$ = 10 nm and averaged gain $b_{w1}$ = 0.2·10$^4$ cm$^{-1}$ or $b_{w2}$ = 0.5·10$^4$ cm$^{-1}$, such that results for high gain of bulk medium can be applied. However, the total thickness of the gain region will be $H_A$~250 nm, as the number of QWs is limited. To ensure mode localization, more than 400 nm thick layers of n- and p-doped semiconductor should be added as well, which increase Im[$k_{eff}$] and decrease device efficiency. Another option is the same as in the QDs system: the uniform distribution of QWs along the waveguide cross section can be applied by increasing the spacing between QWs.

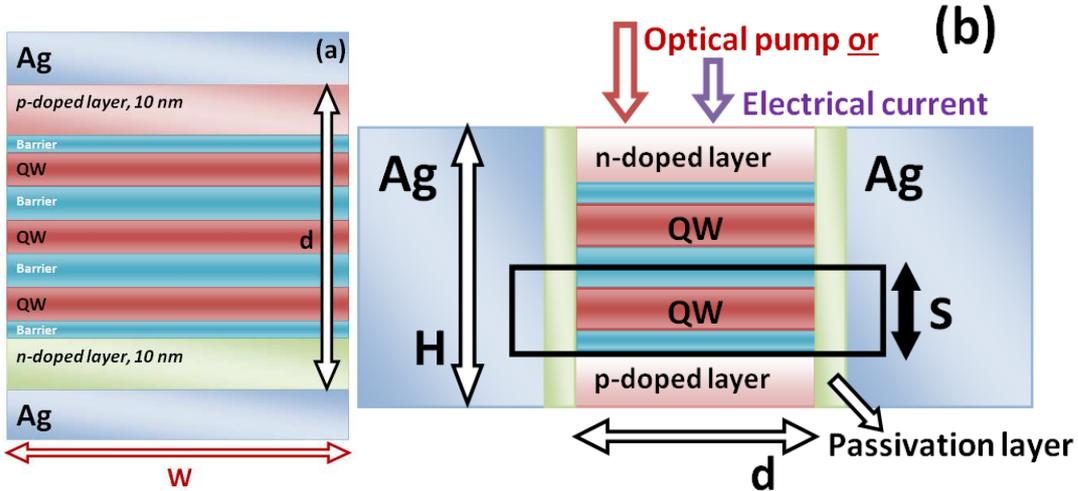

Fig. 9. Cross sections of the MSM waveguide with QWs. The MSM waveguide is composed of In$_{0.466}$Ga$_{0.534}$As quantum wells stack, separated by In$_{0.51}$Ga$_{0.49}$As$_{0.86}$P$_{0.14}$ barriers with n-doped InP layer and p-doped In$_{0.53}$Ga$_{0.47}$As layer. a) Horizontal design. b) Vertical design.

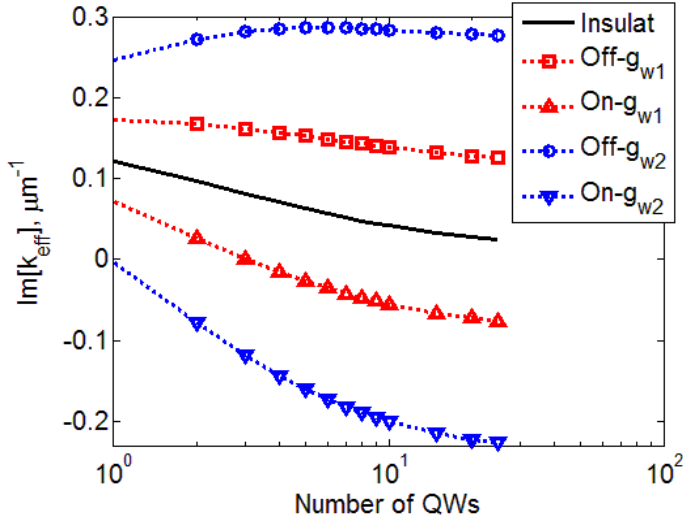

Fig. 10. Imaginary part of the effective propagation constant in the MSM waveguide with the QWs core. Notation: "Insulat" is passive waveguide, $g_{w1}= 0.4 \cdot 10^4$ cm$^{-1}$, $g_{w2}= 1 \cdot 10^4$ cm$^{-1}$.

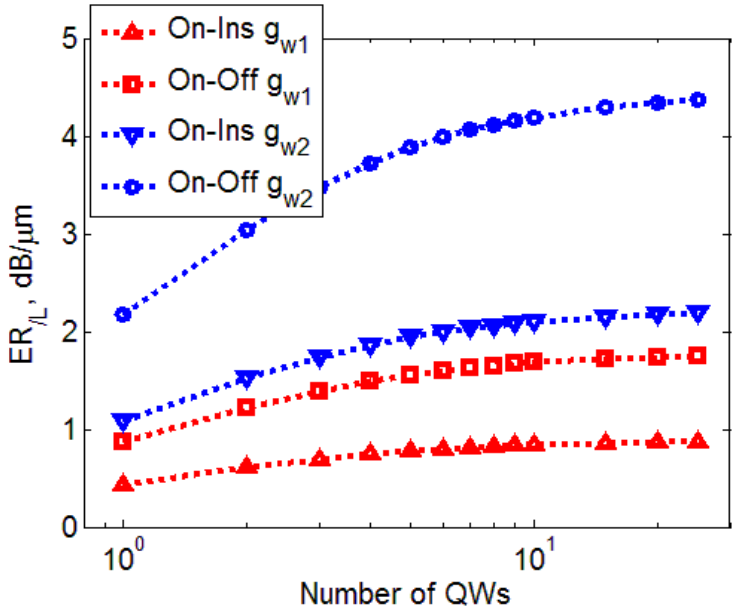

Fig. 11. ER of the MSM device with QWs. "On-Ins $g_{di}$" shows transition between insulator-state ($g = 0$) and on-state with $g_{di}$. "On-Off $g_{di}$" shows transition between off-state with $-g_{di}$ and on-state with $g_{di}$ ($g_{w1}= 0.4 \cdot 10^4$ cm$^{-1}$ and $g_{w2}= 1 \cdot 10^4$ cm$^{-1}$).

## 7. Discussion and conclusions

We have considered three types of plasmonic modulators based on metal-semiconductor-metal waveguides with incorporating gain material. The switching principle is based on suppression of losses in plasmonic waveguides by gain material. The semiconductor core is studied in one of three options: bulk semiconductor, QWs and layers with QDs. In our numerical simulations we considered realistic materials and applied gain parameters achieved in InGaAsP-based structures. We took different gain parameters considered in [37], including such that required very high carrier density. One of the reasons to do this is that we further apply bulk gain medium results for effective approximation of QDs and QWs structures.

We observe promising behaviour in all cases: in principle it is possible not only to compensate losses, but also get waveguide mode amplification. The performance of the proposed modulators is greatly improved in comparison with plasmonic waveguides with controllable losses through electrically driven indium tin oxide layers. In practice, a passive MIM waveguide with $d$ = 120 nm core thickness has Im[$k_{eff}$] = 0.043 μm$^{-1}$. This means 15 dB attenuation on 40 μm length. Following the fabrication requirements we have considered different layouts of the modulators with horizontal or vertical arrangements of gain layers. We can compare horizontal design with bulk semiconductor core and with QWs layers. A MSM waveguide with gain value $g_{b1}$ = $0.1 \cdot 10^4$ cm$^{-1}$ of bulk material and 10 nm n- and p-doped passive layers gives slight amplification, i.e.

Im[$k_{eff}$] = -0.007 μm$^{-1}$. Therefore, the dynamic range of the signal between passive and gain states in the 40 μm-long structure is estimated as 17.5 dB, while between absorbing (Im[$k_{eff}$] = 0.093 μm$^{-1}$) and gain states is doubled to 35 dB. Assuming the same length for a QWs-based MSM waveguide in the horizontal arrangement ($g_{w1}$ = 0.4·10$^4$ cm$^{-1}$ and Im[$k_{eff}$] = - 0.055 μm$^{-1}$) we obtained 34 dB depth of signal modulation between passive and gain states and 52 dB between absorbing and gain states. These results show that the QWs-based waveguides, in the horizontal arrangement are preferable to the bulk based ones. Aside from that, QWs and QDs cores can be compared in vertical arrangement. QWs with $g_{w1}$ = 0.4·10$^4$ cm$^{-1}$ give effective bulk gain $b_{w1}$= 0.2·10$^4$ cm$^{-1}$ of 10 nm thick active layer and QDs with $g_{d1}$ = 1·10$^4$ cm$^{-1}$ give effectively $b_{d1}$= 0.09·10$^4$ cm$^{-1}$ because of small volume. Thus, both structures have approximately the same efficiency.

The FoM (4) of the InGaAsP-based plasmonic modulator is several times higher than the one for MIM plasmonic modulators with the silicon nitride and indium tin oxide multilayered core [15, 35-36]. However, utilizing the studied MSM devices for modulation purposes encounters certain obstacles. The deep modulation requires a full switch of the current. Without enhancement of carrier recombination the response time is in the range of nanoseconds [15]. Nevertheless, modulation with speed up to 10 GHz might be feasible, because of the inherent advantages of the MSM design, i.e. the small dimensions, strong coupling to the plasmonic structure, and enhancement of spontaneous emission due to the tight confinement of modes between two metal plates [51, 52] that serves as electrodes.

Summarizing, all three designs of the MSM core exhibit effective switching. The preference in choosing the optimal design diverts from the pure numerical analysis of the device performance to the question of feasibility of fabrication such metal-semiconductor sandwiches. In addition, decision on vertical or horizontal arrangement of gain layers depends also on the waveguiding parameters, e.g. polarization and coupling efficiency.

Ultra-compact and ultra-fast modulators are among the main requirements for modern photonic integrated circuits. A surface plasmon polariton modulator supplemented with loss-compensation material provides such possibilities. Their potential applications range from direct laser modulation to on-chip optical routing and computation. For this, various in- and out-coupling schemes are needed. While for out-coupling to free space the solution can be straightforward the in-coupling from a VCSEL device is not necessary an easy task. A general in- and out-coupling scheme is difficult to suggest but we believe that, together with the development of plasmonic circuitry in photonics, the coupling problem will be solved.

**Acknowledgments**
R.M. and A.V.L. acknowledges partial financial support from the Danish Research Council for Technology and Production Sciences via the THz COW project.